\documentclass[useAMS,usenatbib,galley]{mn2e}
\def\spose#1{\hbox to 0pt{#1\hss}}
\def\approxlt{\mathrel{\spose{\lower 3pt\hbox{$\sim$}}
	\raise 2.0pt\hbox{$<$}}}
\def\approxgt{\mathrel{\spose{\lower 3pt\hbox{$\sim$}}
	\raise 2.0pt\hbox{$>$}}}
\def\approxpropto{\mathrel{\spose{\lower 3pt\hbox{$\sim$}}
	\raise 2.0pt\hbox{$\propto$}}}
\mathchardef\twiddle="2218

\def\multleft#1{\hbox to size{\vbox {\halign {\lft{##}\cr #1}}\hfill}\par}
\def\multright#1{\hbox to size{\vbox {\halign {\rt{##}\cr #1}}\hfill}\par}

\def\today{\ifcase\month\or January\or February\or March\or April\or May\or
      June\or July\or August\or September\or October\or November\or December\fi
      \space\number\day, \number\year}
\def\<{\thinspace}

\def\arcsec{{\rm\thinspace arcsec}}
\def\cm{{\rm\thinspace cm}}
\def\erg{{\rm\thinspace erg}}

\def\K{{\rm\thinspace K}}
\def\keV{{\rm\thinspace keV}}

\def\km{{\rm\thinspace km}}

\def\Mpc{{\rm\thinspace Mpc}}

\def\s{{\rm\thinspace s}}

\def\ergpcmsqps{\hbox{$\erg\cm^{-2}\s^{-1}\,$}}

\def\ergps{\hbox{$\erg\s^{-1}\,$}}

\usepackage{graphicx}
\usepackage{txfonts}
\usepackage{natbib}
\usepackage{wrapfig}                  
\usepackage{amssymb}
\usepackage{longtable}
\usepackage{times}
\usepackage[usenames]{color}
\usepackage{psfig}
\usepackage{soul,color}
\usepackage{subfigure}
\newcommand*{\mysub}[2]{\ensuremath{#1_{\mathrm{#2}}}}
\newcommand*{\Omegam}{\mysub{\Omega}{m}}

\newcommand*{\Omegal}{\ensuremath{\Omega_{\Lambda}}}

\newcommand*{\LCDM}{\ensuremath{\Lambda}CDM}

\newcommand*{\msolar}{\mysub{M}{\odot}}

\title[The AGN Fraction in Massive Galaxy Clusters]
  { X-ray Bright Active Galactic Nuclei in Massive Galaxy Clusters II: The Fraction of Galaxies Hosting Active Nuclei }
\author[S. Ehlert et al.]
{S.~Ehlert,$^{1,2,3,4}$\thanks{Email:sehlert@space.mit.edu.}
A.~von der Linden,$^{1,3,5}$ 
S.W.~Allen,$^{1,2,3}$
W.N.~Brandt,$^{6,7}$
Y.Q.~Xue,$^{8}$
\newauthor
B.~Luo,$^{6,7}$
A.~Mantz,$^{9,10}$
R.G.~Morris,$^{1,2,3}$
D.~Applegate,$^{1,11}$
and P.~Kelly$^{1,12}$\\
$^1$Kavli Institute for Particle Astrophysics and Cosmology, 452 Lomita Mall, Stanford, CA 94305-4085, USA\\
$^2$SLAC National Accelerator Laboratory, 2575 Sand Hill Road, Menlo Park, CA 94025, USA\\
$^3$Department of Physics, Stanford University, 382 Via Pueblo Mall, Stanford, CA 94305-4060, USA\\
$^4$Kavli Institute for Astrophysics and Space Research, Massachusetts Institute of Technology, 77 Massachusetts Ave., Cambridge, MA 02139, USA\\
$^5$Dark Cosmology Centre, Niels Bohr Institute, University of Copenhagen, Juliane Maries Vej 30, 2100 Copenhagen, Denmark\\
$^6$ Department of Astronomy and Astrophysics, Pennsylvania State University, University Park, PA 16802, USA\\
$^7$ Institute for Gravitation and the Cosmos, Department of Physics, Pennsylvania State University, University Park, PA 16802, USA\\
$^8$ Key Laboratory for Research in Galaxies and Cosmology, Department of Astronomy, \\
University of Science and Technology of China, Chinese Academy of Sciences, Hefei, Anhui 230026, China\\
$^9$ Kavli Institute for Cosmological Physics, 5640 S. Ellis Avenue, Chicago, IL 60637, USA\\
$^{10}$ Department of Astronomy and Astrophysics, University of Chicago, 5640 S. Ellis Avenue, Chicago, IL 60637, USA\\
$^{11}$ Argelander-Institut f\"{u}r Astronomie, Universit\"{a}t Bonn, Auf dem H\"{u}gel 71, 53121 Bonn, Germany\\
$^{12}$ Department of Astronomy, University of California, Berkeley, CA 94720-3411, USA}

\def\cha{{\it Chandra}}
\def\ae{{\small ACIS-EXTRACT}}

\def\sub{{\it Subaru}}
\def\cfht{{\it CFHT}}

\def\rfive{\mysub{r}{500}}
\def\rosat{{\it ROSAT}}

\def\cosm{{\it COSMOS}}

\def\arcsec {\hbox{$^{\prime\prime}$}} 
\def\arcmin {\hbox{$^{\prime}$}} 
\def\MACS {MACS J1931.8-2634}

\def\wav{{\small WAVDETECT}}

\def\cdfs{{\it CDFS}}

\begin{document}

\maketitle

\begin{abstract}
We present a measurement of the fraction of cluster galaxies hosting X-ray bright Active Galactic Nuclei (AGN) as a function of clustercentric distance scaled in units of \rfive. Our analysis employs high quality \cha \ X-ray and \sub \ optical imaging for 42 massive X-ray selected galaxy cluster fields spanning the redshift range of $0.2 < z < 0.7$. In total, our study involves 176 AGN with bright ($R <23$) optical counterparts above a $0.5-8.0 \keV$ flux limit of $10^{-14} \ergpcmsqps$.  When excluding central dominant galaxies from the calculation, we measure a cluster-galaxy AGN fraction in the central regions of the clusters that is $\sim 3 $ times lower that the field value. This fraction increases with clustercentric distance before becoming consistent with the field at $\sim 2.5 \rfive$. Our data exhibit similar radial trends to those observed for star formation and optically selected AGN in cluster member galaxies, both of which are also suppressed near cluster centers to a comparable extent. These results strongly support the idea that X-ray AGN activity and strong star formation are linked through their common dependence on available reservoirs of cold gas.

\end{abstract}

\section{Introduction}

The environments of galaxies play a key role in driving the processes by which galaxies evolve. These effects are expected to be most profound in galaxy clusters, where the density of both the intracluster medium (ICM) and galaxies are highest. Observations have firmly established that galaxy clusters host significantly higher fractions of red, elliptical galaxies than the field \citep[the morphology-density relation, e.g.][]{Dressler1980}. Although significant progress in understanding galaxy evolution has been made, the precise effects of the surrounding intergalactic/intracluster medium (IGM/ICM) and neighboring galaxies on the hot halo gas ($kT \sim 1 \keV$), and cold, star-forming gas ($T \sim 10-100 \K$), in galaxies are still subject to debate.  

Several unique processes are at work in galaxy clusters, each of which influences the hot and cold galaxy gas reservoirs in different ways. Early studies argued that both hot halo gas and cold central gas in cluster galaxies would be stripped by ram pressure on short time scales \citep[$\sim 10-100 \ \rm{Myr}$,][]{Gunn1972,Sarazin1988} compared to the cluster crossing time ($\sim 1-5 \ \rm{Gyr}$). Observational evidence of ram pressure effects has long been observed in cluster galaxies, especially those near the central regions of the nearby Virgo Cluster. Long stripped tails of both hot and cold gas are clearly detected trailing behind Virgo Cluster galaxies \citep{Machacek2004,Oosterloo2005,Kenney2008,Randall2008,EhlertM86}. It was predicted that as the cold gas is removed from cluster galaxies, star formation would be abruptly truncated. However, more recent studies have shown that ram pressure stripping does not strip galaxy gas so effectively \citep[e.g.][]{Larson1980,Balogh2000,Bekki2002,VandenBosch2008a,VonderLinden2010,Wetzel2012}. Rather star formation appears to be quenched on longer time scales of roughly $\sim 1-5 \ \mathrm{Gyr}$. The origin of quenching on this time scale has not been unambiguously determined, although it has been noted that this time scale is similar to the gas depletion time scale for these galaxies \citep[$\mysub{\tau}{gas}=\mysub{M}{gas}/\mathrm{SFR}$, of order $\sim 1 \ \mathrm{Gyr}$, e.g.][]{Larson1980,Bekki2002}. A second class of models, broadly labeled as strangulation, arose to account for these observations and suggested that the ICM does not necessarily strip the majority of gas hosted by cluster galaxies on short time scales. Instead, the cluster environment may prevent the accretion of hotter halo gas onto cluster galaxies \citep{Larson1980,Bekki2002}. In this scenario, star formation rates of cluster galaxies are lower than comparable field galaxies because the existing cold gas reservoirs are depleted by star formation but not replenished by the accretion of ambient halo gas. Such an effect may arise when the hot halo gas in cluster galaxies is efficiently stripped by ram pressure effects while the cold central gas remains bound, although any process that interferes with the replenishment of cold gas reservoirs by accretion can produces a similar outcome. Simulations show that a third process, the repeated tidal interactions between cluster member galaxies (otherwise known as harassment) may also drive key aspects of the morphology-density relationship \citep[e.g.][]{Moore1996,Moore1999}. Harassment has been predicted to bring transport gas towards the centers of cluster galaxies \citep{Moore1999}, and may subsequently result in the triggering of central starbursts in cluster galaxies.   

In contrast to the star formation properties of cluster member galaxies, the occurrence of Active Galactic Nuclei (AGN) in galaxy clusters has not been so well studied. The presence of AGN in galaxies traces mechanisms that transport gas to the galaxy cores, and offer a complementary window onto the unique physical processes that operate on the hot and cold gas reservoirs in cluster galaxies. Early studies of cluster member galaxies \citep[e.g.][]{Osterbrock1960, Gisler1978} determined that optically selected AGN were relatively rare in clusters, occurring less frequently in clusters than in the field. More recent work utilizing large area surveys have expanded these findings, and confirm that the fraction of galaxies hosting optically bright AGN is lower in galaxy clusters than in the field \citep{Kauffmann2004,Popesso2006,VonderLinden2010,Pimbblet2013}. It has been observed, however, when restricting the parent sample to actively star forming galaxies, that the fraction of galaxies hosting strong AGN may not differ in clusters with respect to the field \citep{VonderLinden2010}. Optically faint AGN also appear to occur at similar rates in clusters as in the field \citep[e.g.][]{Martini2002,Best2005,Martini2006,Haggard2010}, while other studies have found no significant differences between optical AGN in dense and sparse environments \citep[e.g.][]{Miller2003}. 

X-ray selected AGN, like optically selected AGN, are typically observed at lower rates in clusters \citep[e.g.][]{Martini2002,Martini2007,Arnold2009,Koulouridis2010,Haines2012}, although the results are limited by the small samples of clusters utilized current studies, typically on the order of $\lesssim 10$. Radio selected AGN are typically observed to be higher in clusters than in the field, especially for Brightest Cluster Galaxies (BCG's) at or near the centers of the cluster potential wells \citep[e.g.][]{Best2004,Best2007,Best2012}. 

These results and others support a consensus view where both optical and X-ray selected AGN are fueled by cold gas ($T \sim 10-100 \K$, sufficiently cold for star formation to occur) while radio-loud AGN may be fueled by the accretion of hotter X-ray emitting halo gas or cold material immediately tied to it \citep{Allen2006,Dunn2008,Dunn2010}. Other studies have provided further evidence for a connection between optical and X-ray AGN. Both X-ray and optically selected AGN are typically observed in massive galaxies \citep[\mysub{M}{\star} $> 10^{10}$ \msolar, e.g.][]{Kauffmann2003,Silverman2009a,Silverman2009b,Xue2010}. The X-ray and optical luminosities of AGN, at least at low redshifts ($z < 1$), correlate with one another \citep{Heckman2005} and their host galaxies tend to have massive, young stellar bulges \citep{Kauffmann2003}. The connection between optical and X-ray AGN does not hold universally, however, as a subset of optically selected AGN with bright line emission are not observed to host X-ray sources \citep[e.g.][]{Heckman2005}. It is also clear from X-ray AGN surveys that the opposite is true, i.e. that many X-ray selected AGN are observed as normal galaxies at optical wavelengths \citep[e.g.][]{Comastri2002,Civano2007,Cocchia2007,Xue2010,Trump2011,Trichas2012}. 

X-ray surveys typically offer the largest projected densities of AGN at any wavelength \citep[see e.g.][for a review]{Brandt2010}. However multiwavelength studies of galaxy clusters, utilizing both X-ray and optical observations, are required to offer a more complete picture of the AGN and galaxy populations within clusters and how the hot and cold gas reservoirs are transformed by the cluster environment. Ram pressure stripping, strangulation, and harassment all offer different predictions for the rate at which galaxies in clusters evolve during infall, a fact that can be exploited to determine the relative contributions of these processes. A galaxy's clustercentric distance is related to the time since it entered the cluster environment. Assuming that central star formation and AGN activity correlate strongly, classical ram pressure stripping \citep{Gunn1972} that would quickly strip hot and cold gas from galaxies would predict an abrupt cutoff in star formation and AGN activity in cluster galaxies at the radius where these ram pressure stripping effects become prevalent. A strangulation-type mechanism will suppress star formation and AGN activity more gradually with infall, while harassment is predicted to trigger central starbursts, and by implication, AGN in cluster galaxies. 

In this paper, we present a determination of the fraction of galaxies hosting X-ray bright AGN (hereafter the X-ray AGN fraction) using a large sample of galaxy clusters with joint X-ray and deep optical imaging data. With these data we are able to, for the first time, measure changes in the X-ray AGN fraction as a function of (scaled) radius to interesting precision, and therefore test the connection between star-forming galaxies and AGN activity in clusters. The structure of this paper is as follows: Section 2 discusses the cluster sample and telescope data used in this study, while Section 3 explains the production of X-ray and optical source catalogs. Section 4 presents the results on the spatially resolved fraction of galaxies hosting X-ray AGN, while Section 5 discusses the physical implications of the data. For calculating distances, we assume a \LCDM \ cosmological model with \Omegam=0.3, \Omegal=0.7, and $H_{0}=70 \km \s^{-1} \Mpc^{-1}$.

\section{The Cluster Sample Selection}
Our study includes a total of 42 galaxy cluster fields (the same sample utilized in \cite{Ehlert2013}, hereafter E13, with the exception of \MACS \footnote{\MACS \ was not included in this study because its location near the Galactic Centre is coincident with a considerably higher stellar density than the other 42 clusters. The field is too crowded to reliably measure aperture magnitudes and separate between stars and galaxies, and any results in this field are therefore not representative of the sample of the other 42 clusters.}). The clusters are drawn from three wide-area, X-ray flux limited cluster surveys derived from the \rosat \ All Sky Survey \citep[RASS;][]{Trumper1993}: the \rosat \ Brightest Cluster Sample \citep[BCS;][]{Ebeling1998}; the \rosat-ESO Flux-Limited X-ray Sample \citep[REFLEX;][]{Bohringer2004}; and the MAssive Cluster Survey \citep[MACS;][]{Ebeling2007,Ebeling2010}. All 42 clusters have been observed with the \cha \ X-ray Observatory. 

Our target clusters are among the most massive and X-ray luminous clusters known, and host large numbers of galaxies and large masses of hot ICM \citep{Mantza,Mantzb}. We therefore expect the influences of the cluster environment to be pronounced in this sample. Robust measurements of cluster masses and the associated characteristic radii, \rfive, are available for all clusters in the sample \citep{Mantza,Mantzb}. With those, we are able to relate observed trends in the AGN population to the virial radii of the clusters. Additionally, each of the target clusters has deep multi-filter optical observations available, taken with the \sub \ and Canada-France-Hawaii ({\it CFHT}) telescopes as part of the ``Weighing the Giants'' project  \citep[][]{VonderLinden2012,Kelly2012,Applegate2012}. These data provide precise measurements of the galaxy populations in these clusters.

\section{Data Preparation}

\subsection{X-Ray Catalog Production}

E13 describe the processes by which the X-ray point sources in the fields were identified, using a procedure similar to the most recent iteration of the \cha \ Deep Field South study \citep[\cdfs,][]{Xue2011}. In short, this procedure starts with an aggressive run of standard \cha \ point-source detection routines optimized for high completeness \citep[\wav,][]{Freeman2002} and follows up each source using the \ae \ analysis package to maximize purity \citep[][]{Broos2010}.\footnote{The \ae \ software package and User's Guide are available at http://www.astro.psu.edu/xray/acis/acis\_analysis.html.} This procedure also allows us to precisely quantify our flux sensitivity to point sources across each cluster field. 

For the purposes of this paper, we further limit our X-ray point source catalogs to those sources that satisfy: 1) A full band ($0.5-8.0 \keV$) flux brighter than $\mysub{F}{X}(0.5-8.0 \keV) > 10^{-14} \ergpcmsqps$; 2) Source positions within $8 \arcmin$ of the exposure weighted mean aimpoint of the \cha \ observations; and 3) an effective exposure time at the source position of at least $20$ ks (see E13 for details). Every source that satisfies these three criteria will have $\geq 15 $ net counts. These criteria were chosen to ensure that the subsample of point sources that satisfy these criteria had nearly $100\%$ completeness and purity. In the \cdfs, the completeness for sources with 15 net counts within $6\arcmin$ of the mean aimpoint has been measured to be $100\%$ and $82\%$ across the entire \cha \ ACIS-I field of view \citep[][]{Xue2011}. Applying these results to our survey, we determine a conservative lower limit on the sample completeness of $90\%$.\footnote{This estimate is conservative in the sense that the average completeness for the \cdfs \ across the entire region beyond $6\arcmin$ from the aimpoint ($\sim 76\%$) is likely significantly lower than the completeness in the radial range of $6-8\arcmin$, given the extent to which the point source sensitivity and point spread function of \cha \ vary with off-axis angle.}  The formal no-source binomial probability (defined and discussed in Paper I) for each X-ray source in this complete subsample is $\sim 10^{-6}$. The number of false detections in this sample is therefore negligible. The flux limit that we choose for this subsample ($\mysub{F}{X}(0.5-8.0 \keV) > 10^{-14} \ergpcmsqps$) is at least a factor of $\sim 3$ higher than than the sensitivity limits for each cluster field, based on the calculations discussed in Paper I.\footnote{Our sensitivity calculation takes into account the variations in the PSF, effective area, vignetting, and background (including the diffuse cluster emission) across each field of view. All of the survey area with at least 20 ks of effective exposure time is sensitive to fluxes well below $\mysub{F}{X}(0.5-8.0 \keV) = 10^{-14} \ergpcmsqps$.}   

In total, 554 X-ray point sources survive all of these cuts. More than $\sim 95\%$ of the X-ray point sources at this flux limit in the \cdfs \ are identified as AGN \citep{Lehmer2012}, and the cluster member point sources can be safely classified as AGN based on their luminosities (see \S \ref{LumCalcs}). The positional uncertainties for our X-ray sources are expected to be no larger than $\sim 1.0 \arcsec$ \citep{Xue2011} throughout the entire region over which we survey. Eight of the galaxy clusters included in E13 have a nominal exposure time less than $20 \mathrm{ks}$, and the X-ray source catalogs for these 8 clusters do not contribute to this work.

\subsection{Optical Catalog Production}

The optical imaging used to identify stars and galaxies in
these cluster fields is described in detail by the {\it
  Weighing the Giants} project, \citet[herafter WtG1]{VonderLinden2012},
\citet[WtG2]{Kelly2012}, and \citet[WtG3]{Applegate2012}. For each cluster, deep imaging taken with SuprimeCam at the \sub \ telescope and/or MegaPrime
at the \cfht \ is available in at least three filters. Here we use the
object catalogs described in Sect.~6.2 of WtG1 - these were produced
with {\tt SExtractor} parameters suitable for identifying larger
objects, such as the galaxies in the clusters described here (in
contrast to the weak lensing catalogs used primarily in the {\it
  Weighing the Giants} project, which require shape measurements of
faint, small galaxies). The optimization for larger objects comes at
the cost of a somewhat higher incompleteness at faint magnitudes;
however, we find that all fields are highly complete at least to
$R<24$ in the SuprimeCam $R$-band. In order to compare the observed AGN fractions to those
inferred from \cosm, we use aperture magnitudes with 3\arcsec\,
diameter. Three of these fields were not observed with the \sub \ $R$-band, and instead we have observations with the MegaPrime {\it r}-filter and {\it g}-filter. We convert the magnitudes into SuprimeCam {\it R}-band using the empirically determined correction formula\footnote{This particular conversion is discussed at http://www.sdss.org/dr5/algorithms/sdssUBVRITransform.html, and we have verified its accuracy on cluster data sets which have been imaged with both \cfht \ and \sub \ filters. }
\begin{equation}
R=r-0.1837 \times (g-r)
\end{equation}
\noindent We distinguish between galaxies (extended
objects) and stars (optical point sources) based on the FWHM and the
{\tt SExtractor} CLASS\_STAR parameter, which are both measured on the detection
image (the image with the best seeing, see WtG1). In 6 fields, the
BCG is saturated in the detection image - in these cases we based the magnitude limits for the catalogs on images in different bands that are not saturated.

\subsection{\cosm \ as a Control Field }

In order to determine the expected properties of the field galaxy population, we utilize X-ray and optical source catalogs from the Cosmic Evolution Survey (\cosm). The X-ray source catalog for the \cha \ \cosm \ survey field was produced using the same procedure used for the cluster fields, while the \cosm \ optical catalog utilized in this study is from \cite{Capak2007}. We utilize the observed magnitudes in all 30 \cosm \ filters to interpolate their \sub \ $R$-band magnitudes, using the photometric redshift calculations discussed in WtG2.  We impose the same cuts for X-ray source fluxes and \cha \ exposure times for the \cosm \ field as we did for the cluster fields.

\subsection{Counterpart Matching}
For our counterpart matching between X-ray and optical source catalogs, we use a fixed $2\arcsec$ matching radius. This matching radius maximizes the number of X-ray sources with optical counterparts while still maintaining an acceptably low rate of expected chance matches. The rate of chance matches was determined by adding offsets to the X-ray source positions and re-running our counterpart matching; we estimate from this calculation that $\sim 10\%$ of the matches may be due to chance coincidence. This matching radius is sufficiently large to account for both positional uncertainties in the X-ray sources due to e.g. the variations in the \cha \ point spread function across the field of view and uncertainties in the overall \cha \ astrometric solutions. As discussed in WtG1, the astrometric solutions for the \sub \ data have absolute precision at the level $\sim 0.1\arcsec$ or better, and the positional uncertainties for our optical data can therefore safely be neglected for this analysis. In the few instances where more than one optical source was located within the matching radius of an X-ray source, the brightest optical source was chosen as the counterpart. Our matching rates for the chosen X-ray and optical flux limits are similar to those determined with deep field surveys such as \cosm, when operating in the same wavebands and flux ranges. 

After performing the matching procedure on the full X-ray and optical catalogs, we select those X-ray sources with optical counterparts that also satisfy the following conditions: (1) the optical counterpart is a galaxy brighter than an $R$-band magnitude of $23$; and (2) the optical counterpart is fainter than $\mysub{R}{BCG}-0.5$, where $\mysub{R}{BCG}$ is the $R$-band magnitude of the BCG for the particular field. We exclude the BCG's themselves from our standard analysis. Because they are subject to different physical processes than typical cluster member galaxies, we will discuss them separately.\footnote{The BCG's in this cluster sample have $R$-band magnitudes in the range of $\mysub{R}{BCG} = 17.7-20.9$, with a median value of $\mysub{R}{BCG}=19.3$.  We caution that since these are aperture magnitudes they will systematically underestimate the actual BCG fluxes, given that most BCG's are larger than $3\arcsec$. The choice of using aperture magnitudes is motivated by the \cosm \ catalogs, for which only aperture magnitudes are public. }  Tests carried out on X-ray sources matched to other optical source populations (discussed in more detail below) show no evidence for cluster member AGN hosted in faint galaxies with $R > 23$ (i.e. excess sources above the expected field counts located in the clusters). Additionally, no cluster member AGN are detected in X-ray sources matched either to optical stars or in X-ray sources without optical counterparts. We therefore conclude that the bulk of the cluster member X-ray AGN are hosted in galaxies with $R < 23$. We apply the same restrictions on the optical counterparts in the \cosm \ control field, using the median BCG magnitude of $\mysub{R}{BCG}-0.5=18.8$ as our upper flux limit. 

\begin{figure}
\centering
\includegraphics[width=0.92\columnwidth, angle=270]{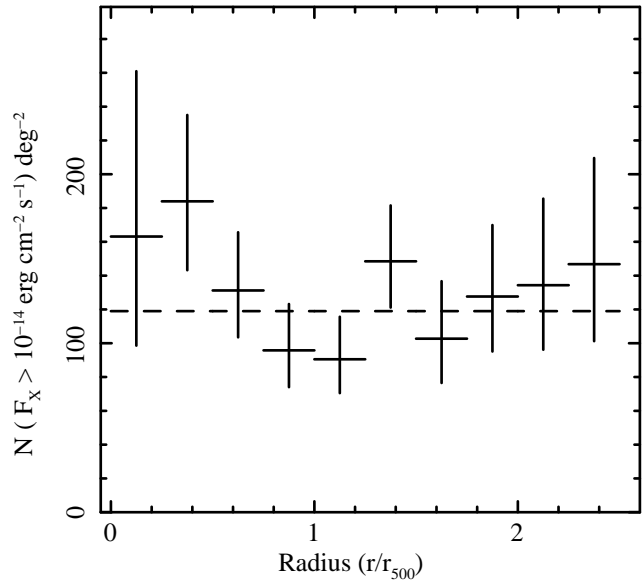}
\caption{\label{XRayProf} The projected number density profile of X-ray point sources with $\mysub{F}{X} (0.5-8.0 \keV) > 10^{-14} \ergpcmsqps$ with optical counterparts located within $2\arcsec$ of the X-ray source position. Radii are scaled in units of \rfive. The magnitude range for the optical counterparts is $\mysub{R}{BCG}-0.5 < R < 23$, where $\mysub{R}{BCG}$ corresponds to the magnitude of the BCG for each individual cluster. Any X-ray point sources matched to BCG's are not included in this profile. The projected source density profile beyond $\sim \rfive$ is consistent with \cosm. The maximum number of X-ray point sources in a radial bin is 29, and the minimum number is 6.  The dashed line denotes the measured source density at this same flux limit for the \cosm \ field.    }

\end{figure}

\subsection{Luminosity and Stellar Mass Limits}\label{LumCalcs}

Our study utilizes flux-limited samples in lieu of absolute magnitude or stellar mass limited samples in order to maximize the sample sizes and hence the statistical power of the study.  Previous studies of the X-ray AGN population in clusters have shown that we expect on average $\sim 1 $ X-ray AGN per cluster field \citep[e.g.][]{Gilmour2009, Ehlert2013} at the X-ray flux limit we utilize. Every step of our analysis is therefore limited by the statistical fluctuations on the X-ray source counts. 

At the optical flux limit discussed above ($ R < 23$), this flux limit corresponds to an absolute magnitude limit of $\mysub{M}{R} \sim -20$ for the highest redshift cluster in this sample (MACS J0744.8+3927, $z=0.697$). This absolute magnitude roughly corresponds to a stellar mass of $\sim 10^{10} \msolar$. Given our overall flux limit, the X-ray AGN sample is complete for all redshifts considered down to a luminosity of $\mysub{L}{X} \sim 1.6 \times 10^{43} \ergps$. For our lowest redshift cluster (Abell 209, $z=0.206$), these same flux limits correspond to $\mysub{M}{R} \sim -17$ and $\mysub{L}{X} \sim 1.1 \times 10^{42} \ergps$. 

Although photometric redshifts
of background galaxies in a subsample of these clusters are presented
in WtG2, identifying galaxies at the cluster redshifts with
photometry is still subject to large uncertainties, in particular for identifying blue or faint galaxies \citep{Rudnick2009}. Therefore any cluster AGN fractions calculated using
photo-z’s would be subject to systematic biases.
To avoid any potential complications that may arise from failing to include cluster member galaxies not on the red sequence in the calculation and provide the largest sample of galaxies (and subsequently most
statistically powerful signal possible), we do not utilize photo-z's here.

\section{Results}

\subsection{The Projected Density of Cluster X-ray Sources} 
We use the procedure discussed in E13 in order to determine
the projected density profile for X-ray point sources. We omit any
X-ray sources located in masked regions of the optical
images, e.g. at the positions of bright stars. Figure \ref{XRayProf} shows the projected density profile of X-ray point sources, $\mysub{N}{X}(\mysub{F}{X} > 10^{-14} \ergpcmsqps, r)$, matched to galaxies with $\mysub{R}{BCG}-0.5 < R < 23$. The total number of X-ray point sources that satisfy all of our previously stated selection criteria is 176.  All of our data points are formally consistent with the field AGN density expected from \cosm, although there is some weak evidence in this curve of a small excess of AGN in the centralmost regions of the clusters. Beyond radii of $\sim 0.5 \rfive$ the source density profile  is consistent with a constant value itself in agreement with expectations from the field (see below). These results are consistent with those of E13, although the stricter selection criteria employed here (especially the optical counterpart matching) reduces the sample size and ultimately the statistical precision which we can measure excess source densities associated with the clusters\footnote{ As compared to the results presented here, the source density profile of Paper I has a lower overall flux limit, no additional restrictions on effective exposure time, and no restrictions with regards to the optical counterparts. In this larger sample a much clearer excess of AGN in the centers of clusters is observed than the results presented here.}.  

Measuring the projected source density profile out beyond $ 1.5 \rfive$ gives a measured field density ($115 \pm 15 \deg^{-2}$) consistent with the expected field density from \cosm \ ($120 \pm 12 \deg^{-2}$). We note that reconstructions of the \cosm \ density field using deep spectroscopic measurements have determined a significant overdensity with respect to the cosmic mean at redshifts of $z \sim 0.8-1.0$ \citep{Kovac2010}, which is also the peak of the redshift distribution for X-ray AGN in field surveys \citep[e.g][]{Brusa2010,Luo2010,Xue2011}. It is therefore possible that the X-ray point source density in the \cosm \ field is systematically slightly higher than the true field source density.

\subsection{The Projected Density of Cluster Galaxies} 
\begin{figure}
\includegraphics[width=0.92\columnwidth, angle=270]{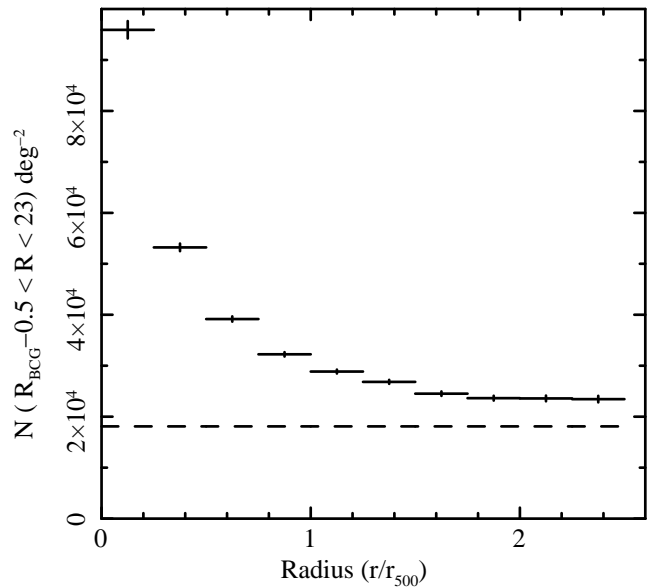}
\caption{\label{OptProf} The projected density of optically selected sources ($ \mysub{R}{BCG}-0.5 < R < 23$) in galaxy cluster fields, as a function of radius in units of \rfive. The dashed line denotes the expected density of field galaxies from \cosm, using the median BCG magnitude of $\mysub{R}{BCG}-0.5=18.83$. A clear excess of galaxies corresponding to a $\sim 15\%$ overdensity above the \cosm \ value is observed at distances of $\sim 2.5 \rfive$.  }
\end{figure}

Figure \ref{OptProf} shows the projected density of optically selected galaxies, \mysub{N}{O}(r), corrected for regions ``masked'' by brighter objects along the line of sight. The total number of bright ($R < 23$) galaxies observed within $8 \arcmin$ of the \cha \ aimpoints is 40,288. A clear excess of sources above the expected field density is observed out to beyond $\sim 2.5 \rfive$. At distances beyond 2.5 \rfive, the source density in the cluster fields is roughly $\sim 15\%$ higher than \cosm, which is unsurprising given that this radius is still within the turnaround radii of the clusters.

\subsection{The Fraction of Cluster Galaxies Hosting X-ray AGN}

\begin{figure}
\centering
\includegraphics[width=0.92\columnwidth, angle=270]{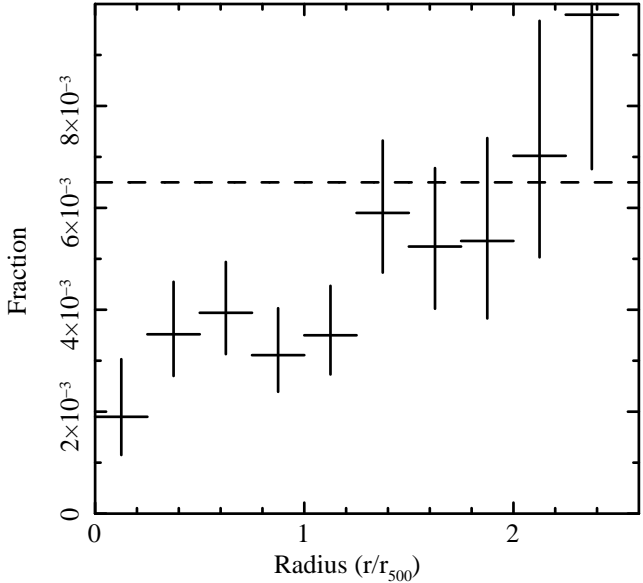}
\caption{\label{AGNFrac} The fraction of cluster+field galaxies ($\mysub{R}{BCG}-0.5 < R < 23$, not including BCG's) hosting X-ray bright ($\mysub{F}{X} > 10^{-14} \ergpcmsqps$) AGN, as a function of radius in units of \rfive. The dashed lines denote the field AGN fraction inferred from \cosm \ at the same limits for the X-ray flux and optical flux, using the median BCG magnitude of $\mysub{R}{BCG,med}-0.5=18.83$. A trend that rises with clustercentric radius is observed, which converges to expected field value at distances of $\sim 2 \rfive$.   }

\end{figure}

Figure \ref{AGNFrac} shows the absolute fraction of cluster+field galaxies hosting X-ray AGN, obtained by counting the total number of X-ray point sources matched to optical galaxies in each radial bin across all of the clusters and dividing by the total number of galaxies in those same radial bins. Because the completeness of the two samples is $\sim 100\%$, this calculation is equivalent to the ratio between the projected source density profiles shown in Figures \ref{XRayProf} and \ref{OptProf}. 

\begin{figure}
\centering
\includegraphics[width=0.92\columnwidth, angle=270]{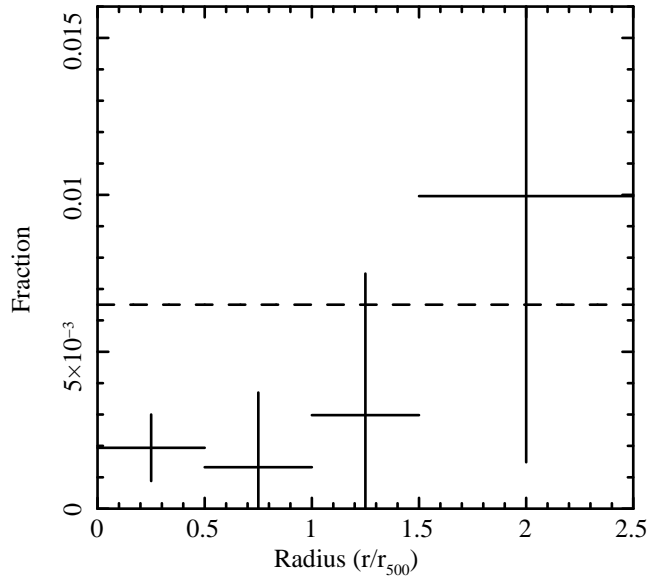}
\caption{\label{AGNFracFieldCorr} The fraction of cluster galaxies ($\mysub{R}{BCG}-0.5 < R < 23$, not including BCG's) hosting X-ray bright ($\mysub{F}{X} > 10^{-14} \ergpcmsqps$) AGN, as a function of radius in units of \rfive. The field population of galaxies and AGN has been subtracted statistically, by fitting the projected source density profiles from Figures \ref{XRayProf} and \ref{OptProf} with constant models, subtracting that constant from both profiles, and dividing them. This shows that the cluster-specific AGN fraction is a factor of $\sim 3$ lower than the field for these same flux limits within $\sim \rfive$.   }

\end{figure}

 The AGN fraction near the cluster centers is clearly lower than the expected \cosm \ value, and slowly rises with radius before converging to a value in statistical agreement with the \cosm \ value at distances of $\gtrsim 1.5 \rfive$. Fitting the AGN fraction between $1.5 $ and $2.5 \rfive$ with a constant model results in a value slightly lower than, but formally consistent with the \cosm \ AGN fraction at these same flux limits given the statistical error bars. A slightly lower AGN fraction in these regions as compared to \cosm \ is not surprising, given that an excess of galaxies is observed all the way out to $\sim 3 \rfive$ while no excess X-ray AGN are observed beyond $\sim \rfive$. In the central-most regions of the clusters (excluding the BCG's), the fraction of cluster+field galaxies hosting X-ray AGN is roughly a factor of 3 lower than the \cosm \ AGN fraction. The AGN fraction shown in Fig. \ref{AGNFrac} is a combination of the cluster AGN fraction and that of the field. In the very central bins, the contribution from background galaxies is modest(see Fig. \ref{OptProf}), but at distances of $r \sim \rfive$, the numbers of cluster and background galaxies are comparable to one another. Using the radial profiles of X-ray sources and galaxies, we have performed a crude statistical subtraction of the field contribution for these radial bins. We used Monte Carlo methods to subtract the field population by fitting the X-ray and optical source density profiles (shown in Figs. 1 and 2, respectively) beyond $2\rfive$ with a constant model. The resulting AGN fraction curve, shown in Figure \ref{AGNFracFieldCorr}, confirms that the cluster AGN fraction is consistently lower than the field fraction within $\sim \rfive$, by a factor of $\sim 3$. The statistical uncertainties on this measurement are too large to constrain the gradient of the cluster-specific AGN fraction, however.

\subsection{The Fraction of BCG's Hosting AGN}

Performing the same analysis procedure and using the same flux limits discussed above, we have also calculated the fraction of BCG's that host X-ray bright AGN. One of the BCG's of the 34 galaxy clusters (MACS J1423.8+2404) in the final sample hosts X-ray bright AGN, for an overall fraction of $2.9\%$. Over the full sample of 43 galaxy clusters, an additional cluster (\MACS) also hosts an X-ray AGN ($4.6\%$). Both of these fractions are consistent with the value measured for the most massive galaxies ($\mysub{M}{\star} \sim 10^{12} \msolar$) hosting X-ray AGN with $\mysub{L}{X} > 10^{42} \ergps$ in the field \citep{Haggard2010}, although with at most two detections this AGN fraction is not well constrained. We emphasize that our selection procedure is highly conservative in identifying point sources in BCG's and only includes the most obvious sources (see E13 for more details). It is therefore possible that our measurement underestimates the true fraction of BCG's hosting X-ray AGN.

\section{Discussion}

Our results provide the best measurements to date of how the fraction of galaxies hosting X-ray AGN varies throughout the cluster environment. Our data shows that the fraction of galaxies hosting X-ray AGN in clusters is consistently lower than the field: the suppression is mild near the edges of the clusters but increases by a factor of $\sim 3$ within $\sim \rfive$.  We now discuss the extent to which these data constrain the physical processes at work in the cluster environment.

First and foremost, these results show no evidence for AGN being triggered in clusters at higher rates than the field. Processes and models that predict higher rates of starbursts and subsequent AGN activity in galaxy clusters with respect to the field \citep[e.g.][]{Moore1996,Moore1999} are therefore disfavored by our data. This places our results in tension with studies that have claimed higher rates of AGN near the viral radii of clusters than the field \citep[e.g.][]{Ruderman2005}. Central starbursts have been observed in a few instances in the outskirts of galaxy clusters \citep[e.g.][]{Moran2005}, and by implication it may be suggested that galaxies near the cluster outskirts may also host higher rates of AGN. For the $\sim 100\%$ complete and pure sample of AGN and galaxies included in this study, however, suppression of AGN appears to already be the dominant process. 

Our results as to how the AGN fraction varies with clustercentric distance are generally consistent with measurements as to how both star formation and optically selected AGN are transformed by the cluster environment. Both star formation and optical AGN are measured to be suppressed to a similar extent as the X-ray AGN results presented here, roughly a factor of $\sim 3$ between the cluster centers and outskirts \citep[e.g.][]{VonderLinden2010,Pimbblet2013}. The fact that X-ray AGN in cluster galaxies are suppressed to a similar degree as star formation and optical AGN activity provides further evidence for a connection between the three processes. Our sample size is not large enough to robustly constrain the time scale over which X-ray AGN activity in cluster galaxies is shut off. Similar studies investigating star formation in cluster galaxies \citep[e.g.][]{VonderLinden2010,Wetzel2012} and optical AGN in clusters \citep[e.g.][]{VonderLinden2010,Pimbblet2013}, however, disfavor scenarios where more efficient versions of ram pressure stripping are operating, and suggest a time scale of $\sim 1 \ \mathrm{Gyr}$. The data presented here are also consistent with such a time scale, and agree with the behavior one expects if AGN activity generally follows central star formation.

A complete model as to how galaxies interact with one another and the ICM that accounts for all of the cluster galaxy observations is still being developed. While our results support the scenario that star formation and AGN in cluster galaxies are being slowly suppressed by the cluster environment, other lines of evidence show that galaxy evolution in clusters is a multi-faceted process. The exact mechanism by which each galaxy is quenched may depend on a number of factors, such as the galaxy's mass and morphology, orbit through the cluster and inclination angle of the disk, and the density and temperature of the cluster ICM. In particular, the precise extent to which halo gas reservoirs may be stripped by ram pressure remains unclear. The slow suppression of star formation is often proposed to originate in the stripping of the hot halo gas from cluster galaxies \citep[e.g.][]{Larson1980,Balogh2000,Bekki2002,VandenBosch2008a}, effectively starving them of gas reservoirs to replenish the cold central gas after it is processed by star formation. However, massive elliptical galaxies in nearby clusters are commonly observed to host bright X-ray halos \citep[e.g.][]{Fabbiano1989}. In other cluster galaxies, the stripping of cold disk gas by ram pressure has also been observed \citep[e.g.][]{Oosterloo2005,Kenney2008,EhlertM86}. Such an observation is not in tension with our results, given that the cold gas stripped from these galaxies likely originates at large distance from the galaxy centers beyond the sphere of influence of the central engine. More complex models should therefore be able to simultaneously describe the survival of the hot gas halo in the most massive galaxy halos, as well as the cold disk stripping in some galaxy/orbit configurations.

The higher fraction of X-ray AGN observed in BCG’s (as compared to other cluster galaxies) is consistent with previous work \citep[e.g.][and references therein]{Hlavacek2013} and demonstrates that BCG’s may be subject to
unique processes that trigger AGN with higher efficiency than typical cluster galaxies. This may be related to the presence of cool
cores in galaxy clusters \citep{Best2007,VonderLinden2007,Fabian2012}; we note that both of the
BCG’s hosting X-ray AGN in this sample are in cool core clusters.
However, not all BCG’s in cool cores (even those undergoing clear
radio-mode AGN feedback) host bright X-ray AGN. A second possible triggering mechanism for BCG’s is the uniquely high rates
of tidal interactions and mergers that occur between the BCG and
other orbiting galaxies. A recent merger between an orbiting galaxy
and the BCG may supply the cold gas to fuel the X-ray AGN. With
only one or two X-ray bright AGN in this sample, however, no robust tests
can be performed that can distinguish between these two scenarios.

Our results provide new and interesting clues regarding the influences of the ICM on AGN, but larger sample sizes of galaxy clusters with joint X-ray and optical observations will be essential to better test and improve our understanding of these processes. In particular, performing a spatially resolved measurement of the X-ray AGN fraction on a stellar mass or absolute magnitude limited galaxy sample will require samples at least a factor of $\sim 3-4$ larger than the sample of 34 presented here. Even larger samples of galaxy clusters will be required to model how the spatially resolved AGN fraction varies with cluster mass or redshift in stellar mass limited galaxy samples. It has been shown that AGN fractions in clusters rise significantly with redshift and ultimately reach levels comparable to or larger than the field at high redshifts \citep[e.g.][]{Hart2009,Martini2009,Hart2011,Martini2013}. While these data cannot offer any new insights into these observed trends, larger samples of clusters will be able to investigate the locations where these new AGN may be triggered in high redshift clusters.   Upcoming wide-field optical surveys will provide deep coverage over large areas of the sky. Therefore the ability to acquire pointed X-ray observations of galaxy clusters in these optical survey fields with \cha \ will ultimately limit the statistical precision with which similar studies can be performed. \cha \ is currently the only available X-ray instrument with the spatial resolution and sensitivity to perform such tests, and will remain so for the foreseeable future. It is therefore critical that it continue to observe large numbers of galaxy clusters and the point sources in those fields so that we can continue to test our understanding of the evolution of galaxies in clusters to new levels of precision.

\section*{Acknowledgments}
Support for this work was provided by the Department of Energy Grant Number DE-AC02-76SF00515 (SE,SWA,RGM) and \cha \ X-ray Center grant GO0-11149X. We also acknowledge support from  NASA ADP grant NNX10AC99G (WNB, YQX, BL) and \cha \ X-ray Center grant SP1-12007A (WNB, YQX, BL), and National Science Foundation grant AST-0838187 (AM). We also acknowledge support from the Thousand Young Talents (QingNianQianRen) program (KJ2030220004; YQX), the USTC startup funding (ZC9850290195; YQX), and the National Natural Science Foundation of China through NSFC-11243008 (YQX). Based in part on data collected at Subaru Telescope (University
of Tokyo) and obtained from the SMOKA, which
is operated by the Astronomy Data Center, National Astronomical
Observatory of Japan. Based on observations obtained
with MegaPrime/MegaCam, a joint project of CFHT and
CEA/DAPNIA, at the Canada-France-Hawaii Telescope (CFHT)
which is operated by the National Research Council (NRC) of
Canada, the Institute National des Sciences de l'Univers of the Centre
National de la Recherche Scientifique of France, and the University
of Hawaii. This research used the facilities of the Canadian
Astronomy Data Centre operated by the National Research Council
of Canada with the support of the Canadian Space Agency.

\bibliographystyle{mnras}
\def \aap {A\&A} 
\def \statisci {Statis. Sci.}
\def \physrep {Phys. Rep.}
\def \pre {Phys.\ Rev.\ E}
\def \sjos {Scand. J. Statis.} 
\def \jrssb {J. Roy. Statist. Soc. B} 


%

\def \araa {ARA\&A}
\def \aj {AJ}
 \def \aas {A\&AS}
  \def \aaps {A\&AS}
\def \apj {ApJ}
\def \apjl {ApJL}
\def \apjs {ApJS}
\def \mnras {MNRAS}
\def \nat {Nat}
 \def \pasp {PASP}
\def \gca {Geochim.\ Cosmochim.\ Acta}
\def \prd {Phys.\ Rev.\ D}
\def \prl {Phys.\ Rev.\ Lett.}

\bibliography{PointSources}

\end{document}